\documentclass[10pt, aps, pra, twocolumn, superscriptaddress]{revtex4-2}

\usepackage{graphicx}
\usepackage{amsmath}
\usepackage{amssymb}
\usepackage{nicefrac}
\usepackage{siunitx}
\usepackage[export]{adjustbox}
\usepackage{ragged2e}
\usepackage{multirow}
\usepackage{xspace}
\usepackage{mathtools}
\setcitestyle{numbers}
\renewcommand{\arraystretch}{2.0}
\DeclarePairedDelimiter\ket{\lvert}{\rangle}
\newcommand{\set}[1]{\{#1\}}
\newcommand{\tuple}[1]{(#1)}

\newcommand{\cells}{\text{\texttt{cells}}}
\newcommand{\colorfunc}{\text{\texttt{color}}}
\newcommand{\colormarkup}{\texttt}

\usepackage[dvipsnames]{xcolor}
\usepackage{hyperref}
\hypersetup{
  colorlinks   = true,
  urlcolor     = MidnightBlue,
  linkcolor    = MidnightBlue,
  citecolor    = MidnightBlue
}
\begin{document}

\title{Decoding 3D color codes with boundaries}

\author{Friederike Butt}
\email[Email to ]{f.butt@fz-juelich.de}
\affiliation{Institute for Theoretical Nanoelectronics (PGI-2), Forschungszentrum J\"{u}lich, J\"{u}lich, Germany}
\affiliation{Institute for Quantum Information, RWTH Aachen University, Aachen, Germany}

\author{Lars Esser}
\affiliation{Institute for Theoretical Nanoelectronics (PGI-2), Forschungszentrum J\"{u}lich, J\"{u}lich, Germany}
\affiliation{Institute for Quantum Information, RWTH Aachen University, Aachen, Germany}

\author{Markus M\"{u}ller}
\affiliation{Institute for Theoretical Nanoelectronics (PGI-2), Forschungszentrum J\"{u}lich, J\"{u}lich, Germany}
\affiliation{Institute for Quantum Information, RWTH Aachen University, Aachen, Germany}

\date{\today}

\begin{abstract}

Practical large-scale quantum computation requires both efficient error correction and robust implementation of logical operations.
Three-dimensional (3D) color codes are a promising candidate for fault-tolerant quantum computation due to their transversal non-Clifford gates, but efficient decoding remains challenging. 
In this work, we extend previous decoders for two-dimensional color codes~\cite{lee2025color}, which are based on the restriction of the decoding problem to a subset of the qubit lattice, to three dimensions. Including boundaries of 3D color codes, we demonstrate that the 3D restriction decoder achieves optimal scaling of the logical error rate and a threshold value of 1.55(6)\% for code-capacity bit- and phase-flip noise, which is almost a factor of two higher than previously reported for this family of codes~\cite{delfosse2014decoding, turner2020decoder}. 
We furthermore present \textsc{qCodePlot3D}, a Python package for visualizing 2D and 3D color codes, error configurations, and decoding paths, which supports the development and analysis of such decoders. 
These advancements contribute to making 3D color codes a more practical option for exploring fault-tolerant quantum computation.

\end{abstract}

\maketitle

\section{Introduction}

Quantum error correction (QEC) encodes information across multiple physical qubits into logical qubits~\cite{gottesman1998theory, aharonov1997fault, knill1998resilient, preskill1998reliable}, such that errors that arise during computation can be detected and corrected. 
In combination with the capability to perform arbitrary and robust operations on encoded qubits, QEC enables the implementation of reliable large-scale computations. 
Recent experiments have demonstrated substantial progress in QEC~\cite{krinner2022realizing, google2025quantum, ryan2024high, ryan2021realization, reichardt2024demonstration, huang2023comparing, postler2023demonstration, nguyen2021demonstration, zhao2022realization, bluvstein2024logical}, as well as fault-tolerant (FT) logical computation on encoded qubits~\cite{pogorelov2025experimental, postler2022demonstration, daguerre2025experimental, lacroix2025scaling, bluvstein2025architectural, gupta2024encoding, chung2025fault, dasu2025breaking}. 
Here, operations are implemented in a way that limits the propagation of errors such that small numbers of errors remain correctable.
Color codes~\cite{Bombin_2006, Bombin_2007} are particularly promising due to their natively transversal gates~\cite{Nielsen_and_Chuang}, where certain logical gates can be 
implemented with only local operations, inherently preventing uncontrolled proliferation of errors. 
Most notably, color codes in three dimensions support a transversal non-Clifford gate, which is a resource that typically requires costly preparation procedures as for example magic-state distillation~\cite{bravyi2012magic}. 
Recently, small instances of three-dimensional (3D) color codes have been realized for the first time~\cite{honciuc2024implementing, daguerre2025experimental, bluvstein2025architectural} to implement FT logical operations, for example by means of code-switching~\cite{butt2024fault, bombin2016dimensional, anderson2014fault}. This approach combines the complementary transversal gate sets of 3D and 2D color codes to implement a universal gate set and has been implemented on trapped ion quantum processors~\cite{daguerre2025experimental, pogorelov2025experimental} as well as neutral atom platforms~\cite{bluvstein2025architectural}.
However, to fully leverage the advantages of color codes, one requires effective methods for implementing QEC by identifying and correcting errors. \textit{Decoding} refers to the task of interpreting error-syndrome measurement outcomes to infer which physical errors have taken place, and to compute a suitable correction that ideally restores the logical state. The accuracy and reliability of this process directly determines the threshold of a code, below which overall error rates are suppressed and long quantum computations become possible~\cite{aliferis2005quantum}.

While surface and toric codes benefit from well-developed decoders with high thresholds, color codes are generally considered to be harder to decode~\cite{takada2024ising}, due to their intrinsic structural properties. 
Existing decoders for 2D color codes are typically based on the restriction of the problem to a subset of the full qubit lattice~\cite{gidney2023new, lee2025color, delfosse2014decoding, turner2020decoder, Chamberland_2020, lee2025color, chamberland2020topological, stephens2014efficient, wang2009graphical, sahay2022decoder, kubica2023efficient}, such that matching-based techniques become applicable. 
Alternative approaches make use of union-find decoding~\cite{bombin2015gauge, delfosse2014decoding}, as well as tensor- or neural- networks~\cite{baireuther2019neural, chubb2021general}, and other strategies~\cite{beni2025tesseract, sarvepalli2012efficient, bombin2012universal}. 
Recently, \textit{vibe} decoding~\cite{koutsioumpas2025colour} as well as neural-network decoding~\cite{senior2025scalable} have achieved 2D color-code performance comparable to that of the surface code in the circuit-level noise setting. 
Restriction-based decoders have been extended to 3D codes without boundaries~\cite{delfosse2014decoding}, but their performance remains below the theoretically predicted optimal threshold and sub-threshold scaling~\cite{kubica2018three, kubica2023efficient, turner2020decoder}. 
In this work, we address this limitation by constructing a restriction-based decoder for color codes with boundaries in three dimensions. Our decoder achieves the optimal sub-threshold scaling and improves on previously reported thresholds for this setting by almost a factor of two~\cite{kubica2023efficient, turner2020decoder}. 

The remainder of this work is structured as follows. In Section~\ref{sec:color_code_construction}, we briefly review the general construction of color codes in three dimensions, including boundaries, and discuss the tetrahedral and cubic color codes. Section~\ref{sec:concatenated_decoder} summarizes Minimum-Weight Perfect Matching and presents our concatenated MWPM Decoder, followed by numerical results in Sec.~\ref{sec:results}. We present \textsc{QCodePlot3D}, which is a python package that we have developed to visualize 3D codes and decoding graphs, in Sec.~\ref{sec:qcodeplot}, and conclude in Sec.~\ref{sec:discussion}.

\begin{figure*}[!tb]
	\centering
	\includegraphics[width=180mm]{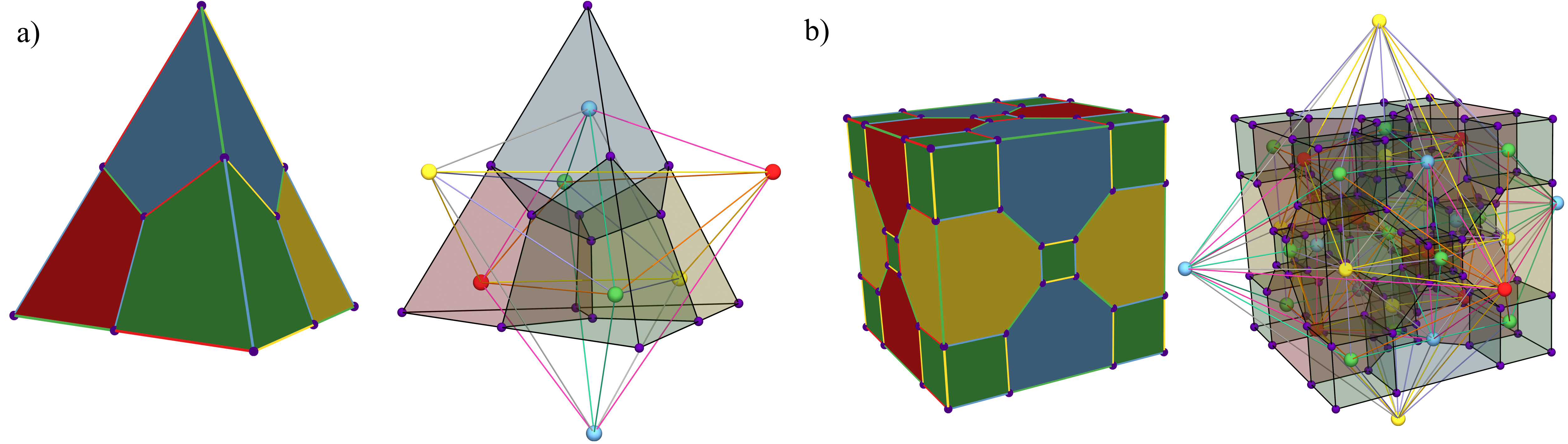}
    \caption{\justifying \textbf{Primal and dual lattices for tetrahedral and cubic color code. } Primal (left) and dual (right) lattice for (a) the distance-3 tetrahedral color code and (b) the distance-4 cubic color code. Vertices of the primal lattice are colored in purple. We sketch the dual lattice on top of the primal one with primal edges depicted in black and primal cells shown with transparent colors for better visualization. Dual vertices and edges have bright colors, the mapping of dual edge colors is: \texttt{rb} is pink, \texttt{rg} is brown, \texttt{ry} is golden, \texttt{bg} is olive green, \texttt{by} is grey and \texttt{gy} is purple. Primal faces and dual cells are not colored for better readability. 
 }
	\label{fig:example_graphs_d_3}
\end{figure*}

\section{Construction of 3D Color Codes without boundaries}\label{sec:color_code_construction}

In this section, we review the formal construction of color codes in three dimensions~\cite{Bombin_2007, bombin2013self, kubica2015universal, bombin2018transversal, kubica2015unfolding} without boundaries,~i.e. the bulk structure of 3D color codes, before discussing the construction with boundaries in the next section. 
3D Color codes can be embedded in a 3D lattice $\mathcal{L} = \tuple{V, E, F, C}$ with vertices $v\in V$, edges $e \in E$, faces $f \in F$ and cells $c \in C$. Here, pairs of vertices $V$ form edges $E$
\begin{equation}
  E \subseteq \set{\set{v_1, v_2} \mid v_1, v_2 \in V \land v_1 \neq v_2}.
\end{equation}
Subsets of vertices form the cells $C$ of the graph, and each cell contains a number of vertices that is a multiple of four
\begin{equation}
  C \subseteq \set{c \mid c \in \mathcal{P}(V) \land |c| = 4k\,\,\mathrm{ for }\,\,k \in \mathbb{N}},
\end{equation}
where $c$ is a connected set of vertices. 
The interfaces of cells correspond to faces $F$ of the graph, which are sets of three or more vertices
\begin{equation}
  F = \set{c_1 \cap c_2 \mid c_1, c_2 \in C \land c_1 \neq c_2}. 
\end{equation}
In the following, we write 
\begin{align}
  \cells(v) &= \set{c \mid v \in c \land c \in C } \\
  \cells(e) &= \set{c \mid e \subset c \land c \in C } \nonumber \\
  \cells(f) &= \set{c \mid f \subset c \land c \in C }\nonumber
\end{align}
for the set of cells that contain vertex $v$, edge $e$ or face $f$. 
One can now assign colors to vertices, edges, faces and cells as a method of bookkeeping~\cite{Bombin_2007}. 
We either assign one \textit{monochrome} color red (\colormarkup{r}), green (\colormarkup{g}), blue (\colormarkup{b}) and yellow (\colormarkup{y}), or a combination of two colors, a mixed color, like \colormarkup{rg} or \colormarkup{by}.

There are two conventionally used definitions of color codes~\cite{dennis2002topological, Bombin_2007, kubica2015unfolding}: placing the qubits either on the lowest-dimensional objects, the vertices, or the highest-dimensional objects, the cells. Both definitions are isomorphic to each other but give rise to different kinds of tessellations with different requirements on the lattice structure, which we discuss in the following. 

The \emph{primal lattice} $\mathcal{L}$ \cite{Bombin_2007} provides a clear visualization of a code. The following two requirements must be fulfilled for a primal graph of a 3D color code:
\begin{enumerate}
    \item Vertices are 4-valent, so each vertex shares an edge with four other vertices.
    \item Cells are 4-colorable, so each cell of $\mathcal{L}$ can be colored with one of the four monochrome colors in such a way that cells sharing a face have different colors.
\end{enumerate}
Additionally, each edge is colored with the monochrome color of the two cells it connects, and each face is colored with the mixed color of the two cells it separates. Now we can use this graph to define a quantum code by placing a qubit at each vertex of $\mathcal{L}$. Each cell $c$  supports an $X$-type stabilizer generator, $S^X_c$, and each face $f$ supports a $Z$-type stabilizer generator $S^Z_f$ 
\begin{align}
  S^X_c &\coloneq \bigotimes_{v \in c} X_v, \label{eq:stab_def_primal}\\
  S^Z_f &\coloneq \bigotimes_{v \in f} Z_v.  \nonumber 
\end{align}

The \emph{dual lattice} $\mathcal{L}^*$~\cite{kubica2015unfolding} is a useful tool for decoding and fulfills the following two criteria:
\begin{enumerate}
    \item Cells are tetrahedra, so each cell has support on four vertices.
    \item Vertices are 4-colorable, so each vertex of $\mathcal{L}^*$ can be colored with one of the four monochrome colors in such a way that vertices sharing an edge have different colors.
\end{enumerate}
Additionally, each face is colored with the monochrome color that is not used by any of its vertices, and each edge is colored with the mixed color of the two vertices it connects. In contrast to the primal lattice, we now place qubits at the cells of $\mathcal{L}^*$. Each vertex $v$ of $\mathcal{L}^*$ supports an $X$-type stabilizer, $S^X_v$, and each edge $e$ supports a $Z$-type stabilizer $S^Z_e$
\begin{align}
  S^X_v &\coloneq \bigotimes_{c \in \cells(v)} X_c \label{eq:stab_def_dual}\\
  S^Z_e &\coloneq \bigotimes_{c \in \cells(e)} Z_c.  \nonumber
\end{align}

In this construction, a vertex in the primal graph is a cell in the dual graph and a face in the primal graph is an edge in the dual graph, and vice versa. Both definitions Eq.~\ref{eq:stab_def_primal} and Eq.~\ref{eq:stab_def_dual} give rise to the same stabilizer definitions.

\begin{figure*}[!tb]
	\centering
	\includegraphics[width=180mm]{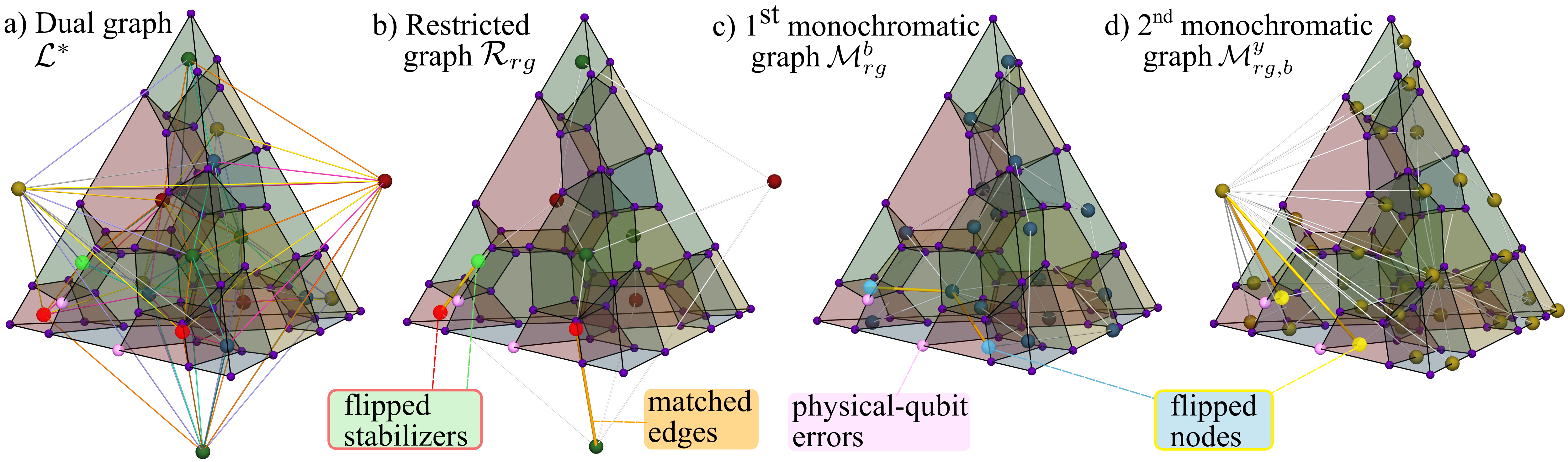}
  \caption{
    \justifying
    \textbf{Example decoding path of the 3D concatenated MWPM decoder on the distance-5 tetrahedral color code.} (a) Dual-graph picture of the distance-5 tetrahedral color code with boundaries, where physical qubits of the primal lattice are shown in purple. Errors are placed at physical qubits shown in pink. For this exemplary error configuration, two neighboring red and one green stabilizers are flipped, as indicated by the bright large nodes on the dual lattice. 
    (b) The restricted graph $\mathcal{R}_{\colormarkup{rg}}$ is constructed by removing all blue and yellow vertices of the dual lattice $\mathcal{L}^*$ and all associated edges. MWPM is run on the restricted lattice and we obtain a set of matched edges (thick orange lines). 
    (c) We obtain the first monochrome graph $\mathcal{M}_{\colormarkup{rg}}^\colormarkup{b}$ by placing blue nodes at all red-green faces as well as all blue cells of the primal graph. Every node that corresponds to an edge of the matching on the restricted graph is marked as flipped (e.g.~the two highlighted light-blue nodes), as well as all initially violated blue nodes. We again run MWPM on this instance yielding a set of matched edges. 
    (d) The second monochrome graph $\mathcal{M}_{\colormarkup{rg,b}}^\colormarkup{y}$ is constructed by placing yellow nodes at all yellow edges and yellow cells of the primal graph. We mark every node that corresponds to a matched edge of the previous step, as well as all initially flipped yellow cells of the initial graph, and again run MWPM. This final matching is the suggested correction of the decoding path.
  }
	\label{fig:decoder/example-decoding}
\end{figure*}

\subsection{3D Color Codes with Boundaries}

The embedding of finite-distance 3D color codes on a 3D lattice requires only locally connected vertices and includes boundaries. Therefore, we need to extend the previous definitions to compact 3D spaces with boundaries.

In the primal picture, there are corner vertices $V_{\mathrm{cor}}$ at the topological boundary that are only 3-valent, instead of 4-valent. 
The number of such 3-valent vertices depends on the tesselation of the lattice and may vary between different codes. 
Analogously, there are edges at the topological lattice boundary that are only part of two faces, instead of three, and faces that are only part of one cell, instead of two. The set of such edges at the topological boundary that connect two corner vertices is called a border.
The sets of faces corresponding to the partition of the topological boundary through the borders are called the graph boundary. 
We modify the requirements on the primal lattice to include boundaries, such that
\begin{enumerate}
    \item Corner vertices are 3-valent, and all other vertices are 4-valent.
    \item Cells in combination with the boundaries are 4-colorable, so each cell and each boundary can be colored with one of the four monochrome colors in such a way that cells and boundaries sharing a face have different colors, and boundaries sharing a border have different colors.
\end{enumerate}
Additionally, the color of a face or edge connecting a cell and a boundary is determined by the color of the respective cell and boundary.
Analogously to faces, borders (i.e.~sets of edges) are colored with the two colors of the two boundaries they separate. 

We analogously adjust the definition of the dual lattice $\mathcal{L}^*$. For each monochrome boundary of $\mathcal{L}$, we add a new vertex of the same color to $\mathcal{L}^*$. Those vertices are the boundaries of $\mathcal{L}^*$.
For each face $f$ between a cell and a boundary, an edge between the respective boundary vertex of $\mathcal{L}^*$ and the vertices of $\mathcal{L}^*$ is added to $\mathcal{L}^*$. An edge is added between two boundary vertices of $\mathcal{L}^*$ if the respective boundaries of $\mathcal{L}$ have a common border.
Additionally, all faces and cells that emerge from the two previous steps are added to $\mathcal{L}^*$. In this extension of the dual lattice, the above requirements are automatically fulfilled. 
The definition of qubits and stabilizers on the graph does not change, except that no $X$-stabilizers are defined at boundaries and no $Z$-stabilizers are defined at borders.

We present two examples of 3D color codes with boundaries in the next subsections to illustrate the construction of the primal and dual lattices as described above. 

\subsubsection*{Tetrahedral 3D Color Codes}

Tetrahedral color codes have four boundaries, as illustrated in Fig.~\ref{fig:example_graphs_d_3}a, and they encode $k=1$ logical qubit. The logical $X$-operator of minimal support is defined on any boundary, the logical $Z$-operator is defined on a border. Table~\ref{tab:3d-cc/color_code_properties_formula} summarizes the number of physical qubits, and independent faces and cells for a given distance $d$. 
The bulk cells of the primal graph are truncated octahedra with six square faces, eight hexagonal faces and 24 vertices, as shown in Fig.~\ref{fig:decoder/example-decoding}.
Cells at the boundary are truncated to either cubes, polyhedra with six square faces, two hexagonal faces and twelve vertices, or polyhedra with six square faces, five hexagonal faces and 18 vertices.

\subsubsection*{Cubic 3D Color Codes}

Cubic color codes have six boundaries, as illustrated in Fig.~\ref{fig:example_graphs_d_3}b. Opposite boundaries share the same color and the code encodes $k=3$ logical qubits. 
The six boundaries of a cubic color code are attributed to three different colors, with two opposing boundaries per color.
We assign each logical qubit to a boundary color \colormarkup{c}. Logical $X$-operators are defined on the \colormarkup{c}-colored boundaries, and the respective logical $Z$-operators on the borders connecting two \colormarkup{c}-colored boundaries. 
The bulk cells of the primal graph are cubes with six square faces and eight vertices, and chamfered cubes \cite{Chamfered_Cubes},~i.e. cubes with symmetrically cut-off edges, with six square faces, twelve hexagonal faces and 32 vertices. Chamfered cubes at the boundary are truncated to either regular cubes, polyhedra with six square faces, seven hexagonal faces and 22 vertices, or polyhedra with seven square faces, eight hexagonal faces, one octagonal face and 28 vertices. Table~\ref{tab:3d-cc/color_code_properties_formula} summarizes the properties of cubic color codes.

\begin{table}[!tb]
    \centering
    \renewcommand*{\arraystretch}{1.7}
    \resizebox{\columnwidth}{!}{%
    \begin{tabular}{|c|c|c|}
    \hline
     & Tetrahedral & Cubic\\
    \hline
    n & $\tfrac{1}{2}(d^3 + d)$ & $5d^3 - 12d^2 + 16$\\
    \hline
    faces & $\tfrac{1}{4}\!\left(\tfrac{5}{3}d^3 - d^2 + \tfrac{7}{3}d - 3\right)$ & $4d^3 - \tfrac{21}{2}d^2 + 3d + 8$\\
    \hline
    cells & $\tfrac{1}{4}\!\left(\tfrac{1}{3}d^3 + d^2 - \tfrac{1}{3}d - 1\right)$ & $d^3 - \tfrac{3}{2}d^2 - 3d + 5$\\
    \hline
    \end{tabular}%
    }
    \caption{\justifying \textbf{Color Code Metrics. }
    Number of physical qubits $n$ and independent graph faces and cells for a given distance $d$ for tetrahedral color codes, encoding $k=1$ logical qubit, and the cubic color code family, encoding $k=3$ logical qubits.}
  \label{tab:3d-cc/color_code_properties_formula}
\end{table}

\section{Decoding 3D color codes with boundaries}\label{sec:concatenated_decoder}

In this section, we briefly review \emph{Minimum Weight Perfect Matching} (MWPM)~\cite{Edmonds_1965} and previous color code decoders that build on MWPM as a subroutine. We then present a new decoder for 3D color codes, the 3D concatenated MWPM decoder, which builds on the concatenated MWPM decoder for 2D color codes introduced in \cite{lee2025color}.

\subsection*{Minimum-Weight Perfect Matching (MWPM)}

MWPM~\cite{Edmonds_1965} is a way of matching up a set of vertices in a graph so that each vertex is matched exactly once. This decoding strategy chooses the pairing with the smallest total edge cost,~i.e. the minimal total edge weight. 
MWPM has been shown to achieve high thresholds for the toric and surface code~\cite{Kitaev_2003, Dennis_2002, higgott2022pymatching, fowler2009high, bravyi2012subsystem}, and to perform efficiently~\cite{fowler2013optimal, paler2023pipelined, higgott2022pymatching},~i.e. with a complexity that scales polynomially with the system size. 
However, MWPM cannot straightforwardly be used to decode color codes, because, generally, single-qubit errors can flip an odd number of stabilizers, which violates the condition to apply MWPM for a perfect matching~\cite{higgott2022pymatching, kubica2023efficient}. 

One way to overcome this problem is to restrict the color code syndrome and to consider only the stabilizers of a subset of colors. On the restricted lattice, a single error is guaranteed to yield an even number of violated stabilizers and is thereby decodable with MWPM. 
The main task of the decoder is then to combine the restricted matchings with a \emph{lifting procedure} to obtain a correction for the original color code~\cite{delfosse2014decoding, kubica2023efficient, turner2020decoder, sahay2022decoder}. Restriction decoders have been used for decoding 2D color codes~\cite{delfosse2014decoding, Chamberland_2020, lee2025color}, where MWPM is run on all three restricted graphs of two colors, and the lifting procedure combines them into a correction of the initial code. In a similar approach~\cite{kubica2023efficient}, MWPM can be run only on two of the three restricted graphs of two colors and a modified \textit{local} lifting procedure is guaranteed to return a global correction. This decoder was generalized to higher-dimensional color codes~\cite{turner2020decoder}, but it does not take boundaries into account and its resulting logical failure rate in leading order does not scale optimally with the physical error rate, but rather as $p_{\mathrm{L}} \propto \mathcal{O}(\frac{d}{3})$. 
Ref.~\cite{lee2025color} describes a decoder for 2D color codes with boundaries, which in turn builds upon the restriction approach and employs a lifting scheme that overcomes the issue of uncorrectable errors of weight $\mathcal{O}(\frac{d}{3})$. In this work, we generalize this decoder~\cite{lee2025color} to 3D color codes.

\begin{figure*}[!tb]
	\centering
	\includegraphics[width=180mm]{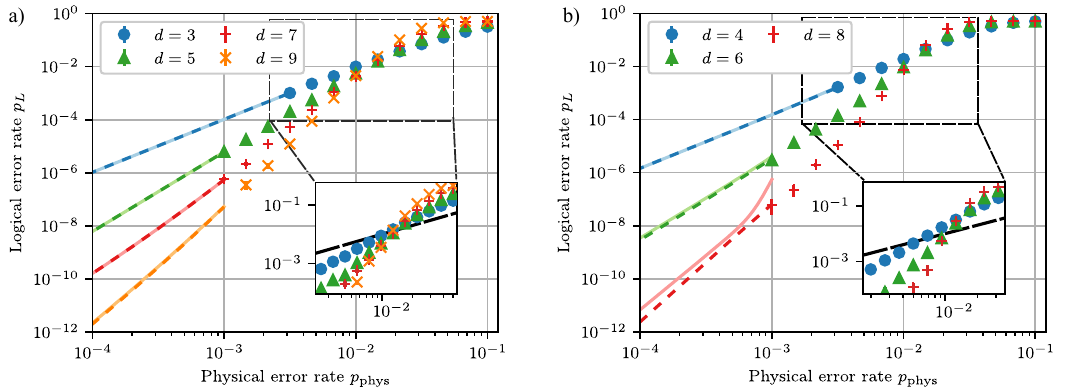}
    \caption{\justifying \textbf{Logical error rates for decoding with the concatenated MWPM decoder. }(a) Decoding of 3D tetrahedral color codes of distance $d = 3, 5, 7, 9$ and (b) $d = 4, 6, 8$ cubic color codes. Data points at physical errors rates $p_{\mathrm{phys}} > 10^{-3}$ are determined by means of direct Monte-Carlo sampling. At low physical error rates, we use Subset Sampling~\cite{Li_2017, heussen2024dynamical} (solid lines) to calculate an upper (light, solid color) and a lower bound (dark, dashed color) on the logical error rate. The inset shows the logical error rates close to (a) $p_{\mathrm{phys}} = 0.014$ and (b) $p_{\mathrm{phys}} = 0.015$, below which increasing distance suppresses the logical error rate. (b) Logical error rates for one of the three encoded logical qubits. The logical error rates of the other two logical qubits of the respective cubic color code are extracted simultaneously and show similar performance, as can be seen in App. Fig.~\ref{fig:app_result_concatenated_tetrahedron}. }
    \label{fig:result_concatenated_tetrahedron}
\end{figure*}

\subsection*{3D Concatenated MWPM Decoder}

In this section, we present our concatenated MWPM decoder for 3D color codes. 
The general idea is to correct errors in 3D color codes by successively applying MWPM on a hierarchy of simplified graphs. 
Each layer captures how errors affect different color subsets of stabilizers. 
By decoding these layers in sequence, the algorithm efficiently reconstructs the most likely set of physical qubit errors from the observed syndromes.

First, we introduce three additional types of graphs: the restricted graph, the first monochrome graph and the second monochrome graph. 
For a dual graph \hbox{$\mathcal{L}^* = \tuple{V_D, E_D, F_D, C_D}$} with vertex colors \colormarkup{c}, \colormarkup{d}, \colormarkup{e} and \colormarkup{f}, we construct the \emph{restricted graph} of colors \colormarkup{c} and \colormarkup{d}, \hbox{$\mathcal{R}_{\colormarkup{cd}} = \tuple{V_R, E_R}$} by removing all vertices of color \colormarkup{e} and \colormarkup{f} and all edges that include vertices of color \colormarkup{e} or \colormarkup{f}
\begin{align}
  V_R &= \set{v \mid \colorfunc(v) \in \set{\colormarkup{c}, \colormarkup{d}} \land v \in V_D} \\
  E_R &= \set{\set{v_1, v_2} \mid v_1, v_2 \in V_R \land v_1 \neq v_2} \nonumber.
\end{align}
For example, the graph $\mathcal{R}_{\colormarkup{rg}}$ contains only red and green vertices and only edges that connect red and green vertices. 
Next, we construct a \emph{first monochrome graph} $\mathcal{M}_\colormarkup{cd}^\colormarkup{e}$, which is based on the previous restricted graph $\mathcal{R}_{\colormarkup{cd}}$ by placing \colormarkup{e}-colored vertices on all edges of $\mathcal{R}_{\colormarkup{cd}}$ (i.e.~\colormarkup{cd}-colored faces of the primal graph) as well as on \colormarkup{e}-colored vertices of $\mathcal{L}^*$. It only contains \colormarkup{e}-colored nodes. 
The monochrome graph of color \colormarkup{e} given the restricted graph $\mathcal{R}_{\colormarkup{cd}}$ is therefore defined as $\mathcal{M}_\colormarkup{cd}^\colormarkup{e} = \tuple{V_{M_1}, E_{M_1}}$
\begin{align}
  V_{M_1} &= E_R \cup \set{\set{v} \mid \colorfunc(v) = \colormarkup{e} \land v \in V_D} \\
  E_{M_1} &= \set{\set{v_1, v_2} \mid v_1 \cup v_2 \in F_D  \land v_1, v_2 \in V_{M_1}}\nonumber
\end{align}
For example, $\mathcal{M}_\colormarkup{rg}^\colormarkup{b}$ contains only blue vertices that are placed on all \colormarkup{rg}-edges and blue vertices of $\mathcal{L}^*$, as shown in Fig.~\ref{fig:decoder/example-decoding}c. Two vertices of the first monochrome graph are connected by an edge iff they correspond to an \colormarkup{f}-colored face of $\mathcal{L}^*$,~i.e. one of them is an edge of $\mathcal{R}_{\colormarkup{cd}}$ and one of them is an \colormarkup{e}-colored vertex of $\mathcal{L}^*$. 

Lastly, we analogously construct the \emph{second monochrome graph} $\mathcal{M}_{\colormarkup{cd},\colormarkup{e}}^\colormarkup{f}$ of color \colormarkup{f} by placing \colormarkup{f}-colored vertices on edges of the first monochrome graph $\mathcal{M}_\colormarkup{cd}^\colormarkup{e}$ (i.e. ~\colormarkup{f}-colored edges of the primal graph) and all \colormarkup{f}-colored vertices of $\mathcal{L}^*$. We define $\mathcal{M}_{\colormarkup{cd},\colormarkup{e}}^\colormarkup{f} = \tuple{V_{M_2}, E_{M_2}}$, based on $\mathcal{M}_\colormarkup{cd}^\colormarkup{e}$, where
\begin{align}
  V_{M_2} &= E_{M_1} \cup \set{\set{v} \mid \colorfunc(v) = \colormarkup{f} \land v \in V_D} \\
  E_{M_2} &= \set{\set{v_1, v_2} \mid v_1 \cup v_2 \in C_D  \land v_1, v_2 \in V_{M_2}}\nonumber
\end{align}
For example, $\mathcal{M}_\colormarkup{rg,b}^\colormarkup{y}$ contains only yellow vertices that are placed on all yellow-edges of $\mathcal{L}$ and yellow vertices of $\mathcal{L}^*$, as illustrated in Fig.~\ref{fig:decoder/example-decoding}d. Two vertices of the second monochrome graph are connected by an edge iff one of them is an edge of $\mathcal{M}_\colormarkup{cd}^\colormarkup{e}$ and one of them is an \colormarkup{f}-colored vertex of $\mathcal{L}^*$.

We now introduce the \emph{3D concatenated MWPM decoder}, first considering $Z$-type errors that are detected by cell-type $X$-stabilizers. For a given code, we can construct the respective dual graph \hbox{$\mathcal{L}^* = \tuple{V_D, E_D, F_D, C_D}$} and track the error syndrome $S \subset V_D$, rooting in a set of $Z$-errors on physical qubits. Figure~\ref{fig:decoder/example-decoding}a shows a distance-5 tetrahedral color code in the dual-lattice picture and an exemplary error configuration. Each error is highlighted in pink and each flipped stabilizer is depicted as a large bright node, while unflipped stabilizers are shown in darker colors.  

In a first step, we construct the restricted graph $\mathcal{R}_{\colormarkup{rg}}$, and mark each node that corresponds to a violated red or green stabilizer, as shown in Fig.~\ref{fig:decoder/example-decoding}b. On the restricted graph, single-qubit errors flip by construction at most two stabilizers, one stabilizer per color since the restricted graph includes only two colors. Therefore, we can use MWPM to match the marked nodes on the restricted lattice (thick orange edges in Fig.~\ref{fig:decoder/example-decoding}b).

Second, we construct the first monochrome graph $\mathcal{M}_{\colormarkup{rg}}^\colormarkup{b}$, and mark nodes that correspond to flipped blue stabilizers or to previously matched edges, as shown in Fig.~\ref{fig:decoder/example-decoding}c. 
Each single-qubit error marks only up to two nodes of the same color:
one node corresponding to a blue stabilizer, and one node corresponding to a previously matched edge.
Therefore, we can again run MWPM to match all marked nodes.

Third, we construct the second monochrome graph $\mathcal{M}_{\colormarkup{rg},\colormarkup{b}}^\colormarkup{y}$, and mark nodes that correspond to flipped yellow stabilizers or to previously matched edges, as shown in Fig.~\ref{fig:decoder/example-decoding}d. We again run MWPM to obtain a final matching, where each edge of the matching corresponds to a physical qubit of the code. The set of qubits obtained from the final matching is the suggested correction of the decoding path \colormarkup{rg,b,y}.

\begin{table*}[!tb]
    \begin{tabular}{|c|c|c|c|c|}
    \hline 
    Code & Pseudo & Cross  & sub-thresh. scaling & effective distance \\
    \hline 
    Tetrahedral CC & 1.13(2)\% & 1.48(2)\% & $p_{\mathrm{phys}}^{(d+1)/2}$ & d\\
    \hline
    Cubic CC & 0.60(6)\% & 1.55(6)\% & $p_{\mathrm{phys}}^{d/2}$ & d-1\\
    \hline 
    \end{tabular}
    \caption{\justifying \textbf{Thresholds for the concatenated MWPM decoder. }We determine the pseudo- and cross-threshold for tetrahedral and cubic color codes, as well as the sub-threshold scaling and the effective distance. These values are extracted from numerical simulations, shown in Fig.~\ref{fig:result_concatenated_tetrahedron}. 
    }
  \label{tab:result_parameters}
\end{table*}

Steps one, two and three are repeated for all other color combinations, as for example for the restricted graph $\mathcal{R}_{\colormarkup{by}}$, the first monochrome graph $\mathcal{M}_{\colormarkup{by}}^\colormarkup{r}$ and the second monochrome graph $\mathcal{M}_{\colormarkup{by,r}}^\colormarkup{g}$.
$\mathcal{L}^*$ has four vertex colors and the restricted graph is created with two colors, so there are $\binom{4}{2} = 6$ restricted graphs.
For each restricted graph, there are two possible combinations to construct the first and second monochrome graph, leading to $6\cdot2 = 12$ possible decoding paths. All 12 decoding paths can be evaluated independently and can be run in parallel. The total runtime is therefore determined by the time it takes to perform one decoding path which includes three MWPM subroutines, as discussed further in App.~\ref{app:runtime_analysis}. 
After evaluating all 12 decoding paths, we select the suggested correction with the lowest weight. 

Our decoder is designed to deal with cell-like stabilizers, and is therefore able to natively correct $Z$-errors. It is also able to correct $X$-errors by combining the face-like stabilizer syndromes to cell-like stabilizer syndromes through multiplying measurement results. 
This merging of face-like stabilizers into cell-like stabilizers reduces the maximum distance $d_x$ for correcting $X$-errors to match the distance $d_z = d$ for $Z$-errors. For example, the tetrahedral $[[15, 1, 3]]$ code, shown in Fig.~\ref{fig:example_graphs_d_3}a, can correct up to weight-three $X$-errors by evaluating all ten face-like $Z$-stabilizers and achieve a distance $d_x = 7$ with a look-up-table decoder. By taking into account only the four $Z$-cells, we reduce this distance to $d_x = d_z = 3$, as we can only correct a single $X$-error. 
More generally, the distance $d_z$ of a 3D color code grows linearly with the length of a border of the 3D lattice which supports the logical operators $Z_{\mathrm{L}}$ for both the tetrahedral and cubic color codes. The distance $d_x$ grows with the size of the $X_{\mathrm{L}}$-operators, which have support on the boundaries, as for example the set of faces on one side of the tetrahedral or cubic lattice. Therefore, $d_x \propto d_z^2$, because the size of a boundary scales quadratically with the length of its borders. In other words, by restricting itself to cell-like stabilizers, the concatenated MWPM decoder induces a square-root reduction in the distance $d_x$, as observed in previous works on 3D color code decoding. The effective distance of the cubic and tetrahedral color codes are summarized in Tab.~\ref{tab:result_parameters}. 
However, the overall distance for an arbitrary input state is always limited by the minimum weight of an arbitrary-error configuration and remains the same. 

In the next section, we use the presented decoding strategy to decode errors on cubic and tetrahedral color codes up to distance nine.

\section{Numerical results}
\label{sec:results}

We numerically simulate the success rate of the presented concatenated MWPM decoder for color codes considering code-capacity noise. Specifically, we prepare perfect logical states $|0\rangle_{\mathrm{L}}$ and $|+\rangle_{\mathrm{L}}$ followed by a noisy idling channel on all data qubits. Here, we apply only bit-flip errors and only phase-flip errors on logical state $|0\rangle_{\mathrm{L}}$ and $|+\rangle_{\mathrm{L}}$, respectively, on each physical qubit with a probability $p_{\mathrm{phys}}$, and then perform noise-free error correction. In the end, we average the total logical failure rate over both logical input states. The logical error rate is determined by means of Monte-Carlo sampling as well as Subset Sampling~\cite{Li_2017, heussen2024dynamical}, as discussed in App.~\ref{app:numerical_methods}. Figure~\ref{fig:result_concatenated_tetrahedron} shows the numerically obtained logical error rates for tetrahedral color codes up to distance $d=9$ and cubic color codes up to a distance $d=8$. 

We determine the \textit{pseudothreshold}~\cite{gupta2024encoding, paetznick2011fault, svore2005flow, chamberland2016thresholds}, which is the breakeven point of the physical error rate and the logical error rate for the lowest-weight error-correcting code, as well as the \textit{cross-threshold}~\cite{google2025quantum, bluvstein2025architectural, paetznick2011fault, svore2005flow, chamberland2016thresholds}, which is the crossing point of the logical failure rates for the same codes at different distances. 
Both of these converge to the \textit{asymptotic threshold}~\cite{paetznick2011fault, svore2005flow, chamberland2016thresholds} for $d \rightarrow \infty$ for an optimal decoder that corrects the theoretically possible maximum number of errors for a given code size. 
We also analyze the scaling of the logical error rate with the physical error rate below threshold, called the \textit{sub-threshold scaling}. 
This scaling determines the effective distance since $p_{\mathrm{L}} \propto p_{\mathrm{phys}} ^{\lfloor (d+1) / 2 \rfloor }$. Table~\ref{tab:result_parameters} summarizes pseudothresholds, cross-thresholds and sub-threshold scalings for the concatenated MWPM decoder run on 3D tetrahedral and cubic color codes. As discussed above, the overall distance is determined by the minimum-weight non-correctable error configuration, and, therefore, fixed by the number of cell-like stabilizer operators. 
Importantly, we find a cross-threshold of 1.48(2)\% for the tetrahedral and 1.55(6)\% for cubic color codes for our concatenated MWPM decoder. 
The ideal asymptotic threshold has been estimated by means of a statistical mechanics mapping to be $p_{\mathrm{3DCC}} \approx 1.9\%$ for string-like logical operators on a general 3D color code~\cite{kubica2018three}. Previous works have reported a cross-threshold of 0.77\% for 3D color codes without boundaries~\cite{kubica2023efficient}, and a cross-threshold of 0.7\% to 0.8\%~\cite{turner2020decoder} for 3D color codes with boundaries. Our decoder therefore improves on these previously reported thresholds by almost a factor of two. 

Furthermore, we can identify a color combination of one decoding path for cubic color codes that achieves the scaling of the logical error rate as reported in Tab.~\ref{tab:result_parameters} without the need for evaluating all 12 decoding paths. 
A single evaluation, and therefore only three instances of MWPM have to be run, which is discussed further in App.~\ref{app:optimal_decoding_path}. 
Here, we effectively localize errors at an early stage in our decoding process by choosing a restricted graph that contains the color green. This choice is motivated by the fact that the set of all green stabilizers has support on every physical qubit. The required instances of MWPM-subroutines can therefore be reduced: before, we needed to perform MWPM on all 6 restricted graphs once, and for each restricted graph there are 2 combinations of monochrome graphs and for each combination two MWPM routines are executed. In total, we can therefore reduce the number of MWPM subroutines from $(1 + 2 + 2) \cdot 6 = 30$ to $3$ for one evaluation of a single decoding path. 
While the scaling of the logical failure rate stays the same as when evaluating all 12 decoding paths, the number of correctable higher-weight error configuration decreases which results in a drop of the pseudothreshold to 0.42(6)\% and the cross-threshold to 1.02(6)\%, as can be seen in Fig.~\ref{fig:individual_paths_d6_cubic}.

\section{Python framework for visualization: \textsc{qCodePlot3D}}\label{sec:qcodeplot}

We have developed a python framework to visualize 2D and 3D color codes and their decoding processes, utilizing the \texttt{Visualization Toolkit} (\texttt{VTK}) \cite{vtkBook} wrapped by the \texttt{pyvista} python package \cite{Sullivan_2019_pyvista}.
It takes a graph $G=\tuple{V, E}$ that describes the color code in the dual-graph picture as an input, and provides an interactive 3D visualization of the color code in the primal-graph picture.
Each family of color codes requires specific postprocessing for optimal layout results.
These are already implemented for tetrahedral and cubic color codes, and can work as guiding examples for visualizing 3D lattice structures. 
Additionally, we provide the functionality to create the dual graph of a cubic/tetrahedral color code for any given even/odd code distance. In both the primal and dual-graph picture, our package offers the possibility to place errors and track their decoding paths. For ease of use, we created a graphical user interface that takes the code parameters of interest as input, and directly constructs the respective interactive graph. 

We distribute \textsc{qCodePlot3D} as a publically available python package \url{https://pypi.org/project/qCodePlot3D}. All figures depicting 3D color codes and their graphs were created with \textsc{qCodePlot3D}.

\section{Discussion}\label{sec:discussion}

In this work, we have developed a new decoder tailored to 3D color codes with boundaries, building on existing decoders for 2D color codes and for 3D codes without boundaries~\cite{lee2025color, kubica2023efficient}. The main contribution of our work is the extension of these decoders~\cite{lee2025color, kubica2023efficient} to three dimensions and to include boundaries, achieving thresholds almost twice as large than previously reported in this setting~\cite{turner2020decoder}. 
This advancement represents an important step toward practical decoding of 3D color codes, which can natively host FT logical operations required for universal quantum computing.
Future work includes the optimization of the presented decoder by exploring trade-offs between runtime and performance. 
Since 3D color code are inherently capable of correcting more $X$-type than $Z$-type errors, adapting the decoder to biased noise models, where one error type occurs more frequently, could further improve the performance. A requirement for this is to use all face-type stabilizers for decoding, not only products of face-operators forming cell-like operators.  
Beyond biased noise, incorporating realistic error sources such as measurement errors, erasure errors and noisy circuit components would bring the decoder closer to the practical deployment in experimental settings~\cite{bluvstein2025architectural, daguerre2025experimental, pogorelov2025experimental}.
In the circuit-level noise setting, additional improvements could be achieved by exploring more sophisticated decoding strategies beyond standard MWPM as a subroutine. One potential approach is to decode individual layers of 2D color codes that collectively form a 3D code instance, in combination with techniques such as \textit{vibe} decoding~\cite{koutsioumpas2025colour} or neural network~\cite{senior2025scalable} decoding, which have shown to perform well under this noise model. 
Overall, our work contributes to the foundation for a range of extensions that could bring 3D color-code decoders closer to practical use in terms of performance and computational efficiency.

\section*{Acknowledgements}
We thank Sascha Heußen for helpful discussions on the layout of tetrahedral color codes and Lukas Bödecker for insightful conversations regarding the weighting of matching graph edges. We gratefully acknowledge support by the European Union’s Horizon Europe research and innovation program under Grant Agreement Number 101114305 (“MILLENION-SGA1” EU Project), 
the German Federal Ministry of Research, Technology and Space (BMFTR) as part of the Research Program Quantum Systems, research project 13N17317 (”SQale”),  
the BMFTR project MUNIQC-ATOMS (Grant No. 13N16070), and the
Intelligence Advanced Research Projects Activity (IARPA), under the Entangled Logical Qubits program through Cooperative Agreement Number W911NF-23-2-0216.
This research is also part of the
Munich Quantum Valley (K-8), which is supported by the
Bavarian state government with funds from the Hightech
Agenda Bayern Plus.
We furthermore acknowledge support by the European ERC Starting Grant QNets
through Grant No. 804247 and 
by the Deutsche Forschungsgemeinschaft (DFG, German Research Foundation) under Germany’s Excellence Strategy ‘Cluster of Excellence Matter and Light for Quantum Computing (ML4Q) EXC 2004/1’ 390534769. The views and conclusions contained in this document are those of the authors and should not be interpreted as representing the official policies, either expressed or implied, of IARPA, the Army Research Office, or the U.S. Government. The U.S. Government is authorized to reproduce and distribute reprints for Government purposes notwithstanding any copyright notation herein. We acknowledge computing time provided at the NHR Center NHR4CES at RWTH Aachen University (Project No. p0020074). This is funded by the Federal Ministry of Education and Research and the state governments participating on the basis of the resolutions of the GWK for national high-performance computing at universities.

\section*{Author contributions:}
F.B. and L.E. developed the presented protocols and L.E. numerically implemented the presented decoders. L.E. developed the python package \textsc{QCodePlot}. L.E. and F.B. performed the numerical simulations and analyzed results. F.B. and L.E. wrote the manuscript, with contributions from all authors. M.M. supervised the project.

\newpage
\appendix

\clearpage

\appendix

\begin{figure*}[!tb]
	\centering
	\includegraphics[width=180mm]{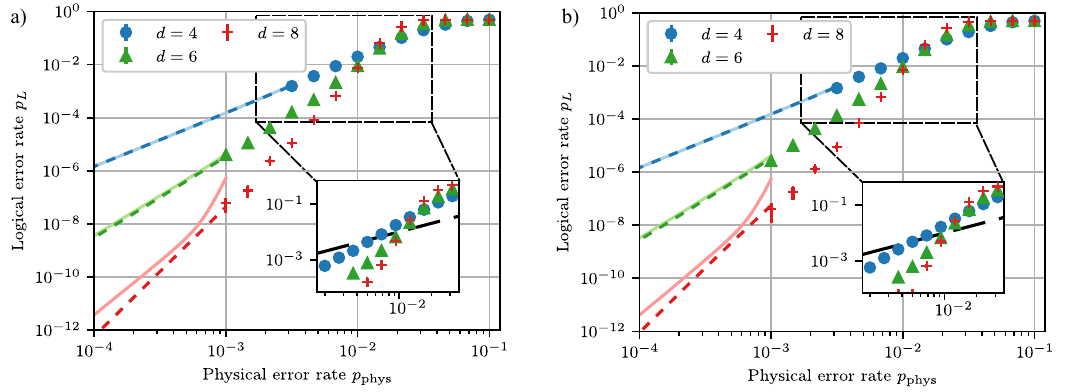}
    \caption{\justifying \textbf{Logical error rates of logical qubits 2 and 3 for decoding cubic color codes with the concatenated MWPM decoder. }(a) Logical error rates for $d = 4, 6, 8$ cubic color codes considering (a) logical qubit 2 and (b) logical qubit 3. Data points at physical error rates $p_{\mathrm{phys}} > 10^{-3}$ are determined by means of direct Monte-Carlo sampling. At low physical error rates, we use Subset Sampling to calculate an upper (light, solid line) and a lower bound (dark, dashed line) on the logical error rate. The inset shows the logical error rates close to $p_{\mathrm{phys}} = $ 1.5\%, below which increasing distance suppresses the logical error rate. All three logical qubits show similar performance. }
    \label{fig:app_result_concatenated_tetrahedron}
\end{figure*}

\section{Numerical Methods}\label{app:numerical_methods}

We numerically determine logical failure rates to evaluate the performance of the presented decoders. We use \emph{direct Monte-Carlo sampling} to simulate logical failure rates for large physical error rates and \emph{Subset Sampling}~\cite{Li_2017, heussen2024dynamical} at small physical errors rates,~i.e. for rare error events. 

\subsection*{Direct Monte-Carlo}

We use the \textsc{Stim} Python-package~\cite{gidney2021stim}, which provides a framework to analyze QEC circuits by means of stabilizer tableau representations~\cite{gidney2021stim}, to perform Monte-Carlo simulations~\cite{Metropolis_1949, Rubino_2009}. The following protocol is implemented numerically: 
\begin{enumerate}
  \item Encode $\ket{0}_{\mathrm{L}}$/$\ket{1}_{\mathrm{L}}$ on each logical qubit $k$ within a QEC code $C$.
  \item Apply an $X$-error to each physical qubit with probability $p_\text{phys}$, following the code-capacity noise model~\cite{Landahl_2011}.
  \item Projectively measure in the $Z$-basis, and use the measurement results to determine the value of the logical operator $Z_{\mathrm{L}}^{(k)}$ and the $Z$-syndrome.
  \item Use a decoder $D$ to decode the syndrome and obtain a Pauli-correction.
  \item Post-process the measurement result of each $Z_{\mathrm{L}}^{(k)}$ by flipping the measurement result for each qubit which is in the support of $Z_{\mathrm{L}}^{(k)}$ and in the correction.
  \item Check if the post-processed measurement result of each $Z_{\mathrm{L}}^{(k)}$ is $+1$/$-1$.
  If this is not the case, a logical error occurred on the respective logical qubit $k$.
\end{enumerate}
For correcting $Z$-errors we use the same protocol for initial states $|+\rangle_{\mathrm{L}}/|-\rangle_{\mathrm{L}}$. By repeating this procedure $m$ times, one can estimate the expectation value of the respective logical operator and logical failure rate for each logical qubit $k$ with
\begin{equation}
  p_{\mathrm{L}}^{(k)}(p_\text{phys}) = \frac{\text{\# logical errors on $k$}}{m}
\end{equation}
The standard deviation of the sampling for large $m$ can be described as~\cite{Rubino_2009}
\begin{equation}
  \label{eq:numerical/mc-standard-deviation}
  \varepsilon_{\mathrm{L}}^{(k)} \sim \sqrt{\frac{p_{\mathrm{L}}^{(k)} (1-p_{\mathrm{L}}^{(k)})}{m}}. 
\end{equation}

\subsection*{Subset Sampling}
In principle, direct Monte-Carlo simulation can be used to determine the logical failure rate for arbitrary small $p_\text{phys}$. However, this becomes inefficient for small $p_\text{phys}$, since actual error events are rare. To achieve results with reasonably small standard deviation, one requires a very large number of repetitions $m$.

A more efficient way to use computational resources for small physical error rates $p_\text{phys}$ is \emph{Subset Sampling}~\cite{Li_2017, heussen2024dynamical}, which samples individual error weights separately.
In the following, we drop the index of the logical qubits $k$  for readability.
For sufficiently small values of $p_\text{phys}$, the most probable setting of a direct Monte-Carlo simulation contains no physical qubit error at all.
Sampling this so called 0-fault subset yields no insights, if it is already known that the decoder works properly for the trivial error configuration.
The next probable setting contains one physical-qubit error, the 1-fault subset, followed by the 2-fault subset with two physical-qubit errors and so on.
One can determine the logical failure rate $p_{\mathrm{L}}^{(\omega)}$ of each $\omega$-fault subset individually, and combine them to the logical failure rate
\begin{equation}
  p_{\mathrm{L}}(p_\text{phys}) = \sum_{\omega=0}^n \binom{n}{\omega} p_\text{phys}^\omega (1-p_\text{phys})^{n-\omega} p_{\mathrm{L}}^{(\omega)}
\end{equation}
The values of $p_{L}^{(\omega)}$ can be sampled individually with $m^{(\omega)}$ repetitions.
One now chooses the maximal cut-off \hbox{$\omega$-fault} subset $\omega_\text{max}$. By considering the cases $p_{L}^{(\omega')} = 0$ and $p_{L}^{(\omega')} = 1$ for each $\omega'>\omega_\text{max}$, respectively, one can estimate a lower and an upper bound for $p_{\mathrm{L}}$.
The lower bound is the sum of the contributions of each subset with $\omega<\omega_\text{max}$, since $p_{L}^{(\omega')} = 0$ lets all terms with $\omega'>\omega_\text{max}$ vanish.
The upper bound contains an additional contribution for $\omega'>\omega_\text{max}$ of
\begin{equation}
  \delta(p_\text{phys}) = 1 - \sum_{\omega=0}^{\omega_\text{max}} \binom{n}{\omega} p_\text{phys}^\omega (1-p_\text{phys})^{n-\omega}
\end{equation}
The standard deviation $\varepsilon_{\mathrm{L}}^{(\omega)}$ of each individual sampling of an $\omega$-fault subset $p_{L}^{(\omega)}$ is described by Eq.~\ref{eq:numerical/mc-standard-deviation}.
The combined standard deviation of all Subset Sampling contributions is
\begin{equation}
  \varepsilon_{\mathrm{L}}(p_\text{phys}) = \sqrt{\sum_{\omega=1}^{\omega_\text{max}} \left[\binom{n}{\omega} p_\text{phys}^\omega (1-p_\text{phys})^{n-\omega} \, \varepsilon_{\mathrm{L}}^{(\omega)}\right]^2}
\end{equation}
The lower and upper bound of the total logical failure rate is given by a combination of the lower and upper bound of $p_{\mathrm{L}}(p_\text{phys})$ with the combined standard deviation of all samplings as
\begin{align}
  p_{\mathrm{L}}^\text{lower} (p_\text{phys})&= p_{\mathrm{L}}(p_\text{phys}) - \varepsilon_{\mathrm{L}}(p_\text{phys}) \\
  p_{\mathrm{L}}^\text{upper}(p_\text{phys}) &= p_{\mathrm{L}}(p_\text{phys}) + \delta(p_\text{phys}) + \varepsilon_{\mathrm{L}}(p_\text{phys}) \nonumber
\end{align}

We do not use \textsc{Stim} to numerically estimate $p_{L}^{(\omega)}$, but determine the measurement results of the logical operators and the stabilizer syndrome directly from the set of physical errors.
If a stabilizer shares an even/odd number of qubits with the set of errors, we would measure $+1$/$-1$.
Given the set of flipped stabilizers, we can run the concatenated MWPM decoder. Since we know the initially prepared logical state, we can track the initially placed error and determine if the respective logical state has been flipped in the end. 

\subsection*{Minimum-Wight Perfect Matching}

We use \textsc{PyMatching} for each MWPM subroutine~\cite{higgott2022pymatching}, which is an open-source package in python available at \url{https://github.com/oscarhiggott/PyMatching}.

\section{Explicit code parameters for $d < 10$ color codes}\label{app:explicit_code_params}

Table~\ref{tab:3d-cc/color_code_properties} shows the number of physical qubits, as well as the number of independent faces and cells of the tetrahedral and cubic color code for distances $d<10$. These were determined by constructing each graph explicitly and then counting the respective quantity. Table~\ref{tab:3d-cc/color_code_properties_formula} summarizes the analytical expressions for these quantities as a function of the code distance $d$.  

\begin{table}[!tb]
    \centering
    \renewcommand*{\arraystretch}{1.7}
    \begin{tabular}{|lccc|}
    \hline
    (a) $d$ & $n$ & \#Z-generators & \#X-generators \\
    \hline
    3 & 15 & 10 & 4 \\ 
    5 & 65 & 48 & 16 \\
    7 & 175 & 134 & 40 \\
    9 & 369 & 288 & 80 \\
    \hline
    (b) $d$ &$n$ & \#Z-generators & \#X-generators \\
    2 & 8 & 4 & 1 \\
    4 & 144 & 108 & 33 \\
    6 & 664  & 512 & 149  \\
    8 & 1808 & 1408 & 397 \\
    \hline
    \end{tabular}
    \caption{\justifying \textbf{Color Code Metrics. }
    Distance $d$, number of physical qubits $n$, independent $Z$-generators (defined on faces) and independent $X$-generators (defined on cells) of the first four members of the (a) tetrahedral color code, encoding $k=1$ logical qubit, and (b) the cubic color code family, encoding $k=3$ logical qubits.
  }
  \label{tab:3d-cc/color_code_properties}
\end{table}

\section{Runtime analysis}\label{app:runtime_analysis}

\begin{figure}[!tb]
	\centering
	\includegraphics[width=88mm]{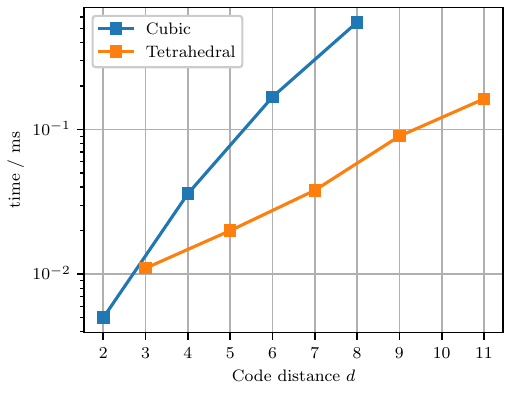}
    \caption{\justifying \textbf{Runtime of the concatenated decoder. }We determine the simulation runtime of one decoding path on a single Laptop (Apple M1 Pro) of the concatenated MWPM decoder for the cubic (blue) and the tetrahedral color code (orange). This includes the three MWPM subroutines and the time it takes to generate the syndrome graphs for each subroutine. }
    \label{fig:runtime_analysis}
\end{figure}

Figure~\ref{fig:runtime_analysis} shows the scaling of the runtime of the numerical simulation for one decoding path, which contains three MWPM subroutines. Decoding the cubic color code at a given distance $d$ takes longer than decoding the tetrahedral of the next closest distance $d+1$, because the number of cells, faces and edges is substantially larger in the cubic color code lattice. In the restricted and monochrome graphs, cells, faces as well as edges correspond to nodes in the decoding graph, and the decoding complexity increases with the number of nodes to match. For example, the $d=4$ cubic code contains 33 cells, 108 faces and 144 vertices while the $d=5$ tetrahedral code only includes 16 cells, 48 faces and 65 vertices, as specified in Tab.~\ref{tab:3d-cc/color_code_properties_formula} and Tab.~\ref{tab:3d-cc/color_code_properties}. 

While Fig.~\ref{fig:runtime_analysis} shows the numerically determined runtime for evaluating a full decoding path, we can additionally estimate the expected runtime of MWPM within our decoding protocol by neglecting finite-size effects as well as the additional overhead associated with constructing the syndrome graphs.
\textsc{PyMatching} uses the \textit{blossom} algorithm~\cite{higgott2022pymatching} for decoding on the complete syndrome graph. For large lattices, its runtime scales with $\mathcal{O}(|\textbf{s}|^3 \log|\textbf{s}|)$, where $\textbf{s}$ is the syndrome vector and $|\textbf{s}|$ is the length of the syndrome vector. One decoding path of the concatenated MWPM decoder includes three matching subroutines. The first subroutine performs matching on the restricted graph, where the length of the syndrome vector scales with the number of cells $|\textbf{s}_1| \propto \#\mathrm{cells}$. The second subroutine performs MWPM on the first monochromatic graph, which includes nodes on faces and cells, meaning $|\textbf{s}_2| \propto \#\mathrm{cells} + \# \mathrm{faces}$. Analogously, the third subroutine matches nodes placed on edges and cells and therefore $|\textbf{s}_3| \propto \#\mathrm{cells} + \# \mathrm{edges}$. Since the number of cells, faces and edges scale with $d^3$, as summarized in Tab.~\ref{tab:3d-cc/color_code_properties_formula}, the total runtime of the MWPM matching decoder scales with $\mathcal{O}(d^9 \log(d))$. 

\section{Logical error rate of logical qubits in a cubic color code}

Figure~\ref{fig:result_concatenated_tetrahedron} shows the logical error rates obtained for logical qubit 1 encoded in a $d= 4, 6, 8$ cubic color code. We achieve a similar pseudo- and cross-thresholds for the remaining two encoded logical qubits, as shown in Fig.~\ref{fig:app_result_concatenated_tetrahedron}.

\begin{figure*}[!tb]
	\centering
	\includegraphics[width=180mm]{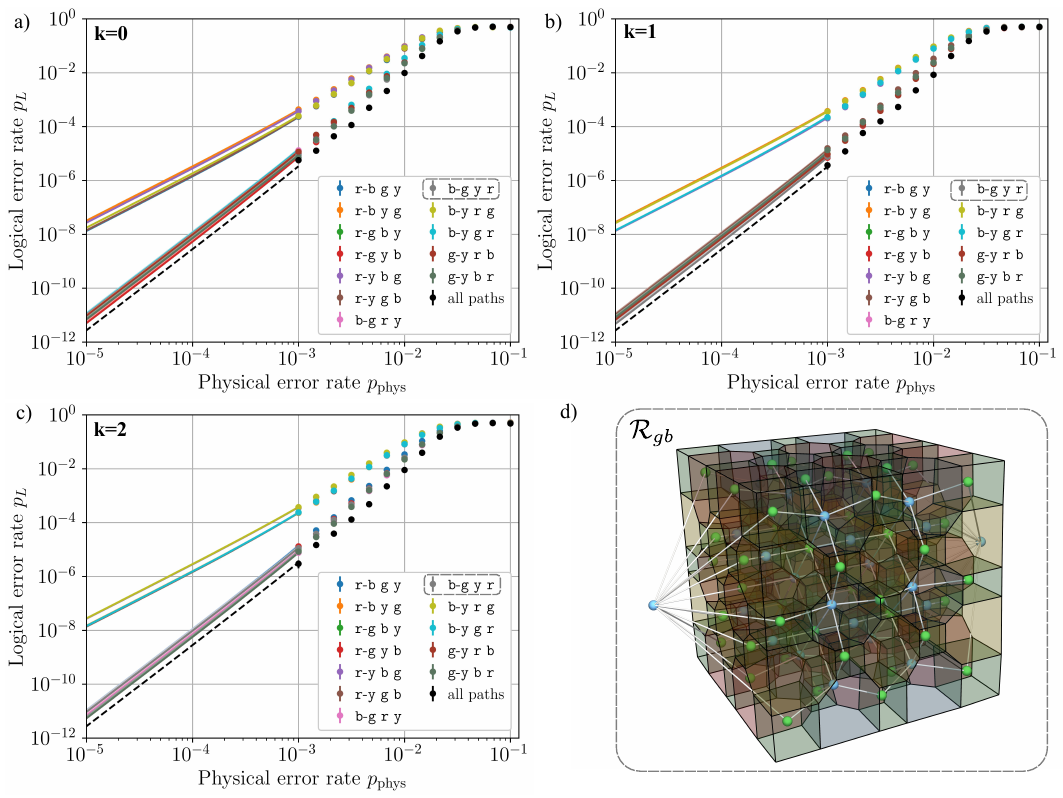}
    \caption{\justifying \textbf{Individual decoding paths for the $d=6$ cubic color code. } Logical error rates evaluated for all 12 decoding paths for logical qubit (a) 0, (b) 1 and (c) 2. There are single decoding paths, for which the logical error rate scales with \hbox{$p_{\mathrm{L}}\propto p_{\mathrm{phys}}^3$} with the physical error rate for all three logical qubits, as for example the \colormarkup{b-g,y,r} path. This means there exists a single decoding path that can be used to correct any weight-2 error and the effective distance is the same as indicated in Tab.~\ref{tab:result_parameters}. (d) \colormarkup{b-g} restricted lattice. The green cells in combination have support on all physical qubits. }\label{fig:individual_paths_d6_cubic}
\end{figure*}

\section{Optimal decoding path for cubic color codes}\label{app:optimal_decoding_path}

Our goal is to reduce the number of decoding paths that have to be evaluated to relax the requirements on computational resources. If decoding paths cannot be evaluated in parallel, fewer decoding paths correspond to shorter runtime since the number of MWPM-subroutines can be decreased. 
We find that a single decoding path suffices to achieve the optimal sub-threshold scaling, considering all three logical qubits in a cubic color code. 
As an example, Fig.~\ref{fig:individual_paths_d6_cubic} shows all 12 individual decoding paths for each logical qubit $k = 0, 1, 2$ for a distance-6 cubic color code. We can identify single paths, as for example \colormarkup{(bg,y,r)} or \colormarkup{(rg,y,b)}, that lead to the desired scaling of $p_{\mathrm{phys}} \propto p^3$ for all three logical qubits. 
Here, all optimal individual decoding paths contain the color green in their restricted graph, which can be attributed to the fact that the joint support of all green stabilizers entails all physical qubits, while it does not for the three other color \colormarkup{r, b, y}. 
This means that we effectively localize errors on any physical qubit at an early stage of our decoding process by including green in the first restricted graph. 
We can always choose an optimal decoding path and therefore reduce the number of required MWPM-subroutines. Before, we had to consider two combinations of MWPM on two subsequent monochrome graphs for each of the 6 restricted graphs, so in total $(1 + 2 + 2 ) \cdot 6 = 30$ MWPM-instances. For a single decoding path, this can be reduced to $1 + 1 + 1 = 3$. However, this decreases the pseudothreshold from 0.60(6)\% to 0.42(6)\% and the cross-threshold from 1.55(6)\% to 1.02(6)\%. 
We did not find an optimal single decoding path for the tetrahedral code, as there is no set of single-color stabilizers that has support on all physical qubits. 

\clearpage
\bibliography{references}

@book{Nielsen_and_Chuang,
   place={Cambridge},
   title={Quantum Computation and Quantum Information: 10th Anniversary Edition},
   DOI={10.1017/CBO9780511976667},
   publisher={Cambridge University Press},
   author={Nielsen, Michael A. and Chuang, Isaac L.},
   year={2010}
}

@article{knill1998resilient,
  title={Resilient quantum computation},
  author={Knill, Emanuel and Laflamme, Raymond and Zurek, Wojciech H},
  journal={Science},
  volume={279},
  number={5349},
  pages={342--345},
  year={1998},
  publisher={American Association for the Advancement of Science},
  doi={10.1126/science.279.5349.342}
}

@inproceedings{aharonov1997fault,
  title={Fault-tolerant quantum computation with constant error},
  author={Aharonov, Dorit and Ben-Or, Michael},
  booktitle={Proceedings of the twenty-ninth annual ACM symposium on Theory of computing},
  pages={176--188},
  year={1997},
  url={https://dl.acm.org/doi/pdf/10.1145/258533.258579}
}

@article{gottesman1998theory,
  title={Theory of fault-tolerant quantum computation},
  author={Gottesman, Daniel},
  journal={Phys. Rev. A},
  volume={57},
  number={1},
  pages={127},
  year={1998},
  publisher={APS},
  doi={10.1103/PhysRevA.57.127}
}

@article{preskill1998reliable,
  title={Reliable quantum computers},
  author={Preskill, John},
  journal={Proceedings of the Royal Society of London},
  volume={454},
  number={1969},
  pages={385--410},
  year={1998},
  publisher={The Royal Society},
  doi={10.1098/rspa.1998.0167}
}

@article{ryan2024high,
  title={High-fidelity teleportation of a logical qubit using transversal gates and lattice surgery},
  author={Ryan-Anderson, C and Brown, NC and Baldwin, CH and Dreiling, JM and Foltz, C and Gaebler, JP and Gatterman, TM and Hewitt, N and Holliman, C and Horst, CV and others},
  journal={Science},
  volume={385},
  number={6715},
  pages={1327--1331},
  year={2024},
  publisher={American Association for the Advancement of Science},
  doi={10.1126/science.adp6016}
}

@article{google2025quantum,
  title={Quantum error correction below the surface code threshold},
  volume={638},
  number={8052},
  pages={920--926},
  year={2025},
  publisher={Nature Publishing Group UK London},
  doi={10.1038/s41586-024-08449-y},
  journal={Nature}
}

@article{krinner2022realizing,
  title={Realizing repeated quantum error correction in a distance-three surface code},
  author={Krinner, Sebastian and Lacroix, Nathan and Remm, Ants and Di Paolo, Agustin and Genois, Elie and Leroux, Catherine and Hellings, Christoph and Lazar, Stefania and Swiadek, Francois and Herrmann, Johannes and others},
  journal={Nature},
  volume={605},
  pages={669--674},
  year={2022},
  publisher={Nature Publishing Group UK London},
  doi={10.1038/s41586-022-04566-8}
}

@article{ryan2021realization,
  title={Realization of real-time fault-tolerant quantum error correction},
  author={Ryan-Anderson, Ciaran and Bohnet, Justin G and Lee, Kenny and Gresh, Daniel and Hankin, Aaron and Gaebler, JP and Francois, David and Chernoguzov, Alexander and Lucchetti, Dominic and Brown, Natalie C and others},
  journal={Phys. Rev. X},
  volume={11},
  number={4},
  pages={041058},
  year={2021},
  publisher={APS},
  doi={10.1103/PhysRevX.11.041058}
}

@misc{reichardt2024demonstration,
  title={Demonstration of quantum computation and error correction with a tesseract code},
  author={Reichardt, Ben W and Aasen, David and Chao, Rui and Chernoguzov, Alex and van Dam, Wim and Gaebler, John P and Gresh, Dan and Lucchetti, Dominic and Mills, Michael and Moses, Steven A and others},
  year={2024},
  note = {{P}reprint at arXiv:2409.04628 },
  url = {https://arxiv.org/abs/2409.04628}
}

@article{postler2023demonstration,
  title = {Demonstration of Fault-Tolerant {S}teane Quantum Error Correction},
  author = {Postler, Lukas and Butt, Friederike and Pogorelov, Ivan and Marciniak, Christian D. and Heu\ss{}en, Sascha and Blatt, Rainer and Schindler, Philipp and Rispler, Manuel and M\"uller, Markus and Monz, Thomas},
  journal = {PRX Quantum},
  volume = {5},
  issue = {3},
  pages = {030326},
  numpages = {19},
  year = {2024},
  month = {Aug},
  publisher = {American Physical Society},
  doi = {10.1103/PRXQuantum.5.030326},
  url= {https://link.aps.org/doi/10.1103/PRXQuantum.5.030326}
}

@article{huang2023comparing,
  title={Comparing {S}hor and {S}teane error correction using the {B}acon-{S}hor code},
  author={Huang, Shilin and Brown, Kenneth R and Cetina, Marko},
  journal={Science Advances},
  volume={10},
  number={45},
  pages={eadp2008},
  year={2024},
  publisher={American Association for the Advancement of Science},
  doi={10.1126/sciadv.adp2008}
}

@article{zhao2022realization,
  title={Realization of an error-correcting surface code with superconducting qubits},
  author={Zhao, Youwei and Ye, Yangsen and Huang, He-Liang and Zhang, Yiming and Wu, Dachao and Guan, Huijie and Zhu, Qingling and Wei, Zuolin and He, Tan and Cao, Sirui and others},
  journal={Phys. Rev. Lett.},
  volume={129},
  number={3},
  pages={030501},
  year={2022},
  publisher={APS},
  doi={10.1103/PhysRevLett.129.030501}
}

@article{bluvstein2024logical,
  title={Logical quantum processor based on reconfigurable atom arrays},
  author={Bluvstein, Dolev and Evered, Simon J and Geim, Alexandra A and Li, Sophie H and Zhou, Hengyun and Manovitz, Tom and Ebadi, Sepehr and Cain, Madelyn and Kalinowski, Marcin and Hangleiter, Dominik and others},
  journal={Nature},
  volume={626},
  pages={58--65},
  year={2024},
  publisher={Nature Publishing Group UK London},
  doi={10.1038/s41586-023-06927-3}
}

@article{chung2025fault,
  title={Fault-tolerant operation and materials science with neutral atom logical qubits},
  author={Chung, Woo Chang and Cole, Daniel C and Gokhale, Pranav and Jones, Eric B and Kuper, Kevin W and Mason, David and Omole, Victory and Radnaev, Alexander G and Rines, Rich and Teo, Mariesa H and others},
  journal={npj Quantum Information},
  year={2025},
  publisher={Nature Publishing Group UK London},
  url={https://doi.org/10.1038/s41534-025-01095-w}
}

@article{takada2024ising,
  title={Ising model formulation for highly accurate topological color codes decoding},
  author={Takada, Yugo and Takeuchi, Yusaku and Fujii, Keisuke},
  journal={Phys. Rev. Res.},
  volume={6},
  number={1},
  pages={013092},
  year={2024},
  publisher={APS},
  doi={10.1103/PhysRevResearch.6.013092}
}

@article{sahay2022decoder,
  title={Decoder for the triangular color code by matching on a {M}{\"o}bius strip},
  author={Sahay, Kaavya and Brown, Benjamin J},
  journal={PRX Quantum},
  volume={3},
  number={1},
  pages={010310},
  year={2022},
  publisher={APS},
  doi={10.1103/PRXQuantum.3.010310}
}

@misc{chubb2021general,
  title={General tensor network decoding of 2{D} {P}auli codes},
  author={Chubb, Christopher T},
  note = {{P}reprint at arXiv:2101.04125},
  url={https://doi.org/10.48550/arXiv.2101.04125},
  year={2021}
}

@article{bombin2015gauge,
  title={Gauge color codes: optimal transversal gates and gauge fixing in topological stabilizer codes},
  author={Bomb{\'\i}n, H{\'e}ctor},
  journal={New J. Phys.},
  volume={17},
  number={8},
  pages={083002},
  year={2015},
  publisher={IOP Publishing},
  doi={10.1088/1367-2630/17/8/083002}
}

@article{bombin2012universal,
  title={Universal topological phase of two-dimensional stabilizer codes},
  author={Bombin, Hector and Duclos-Cianci, Guillaume and Poulin, David},
  journal={New J. Phys.},
  volume={14},
  number={7},
  pages={073048},
  year={2012},
  publisher={IOP Publishing},
  doi={10.1088/1367-2630/14/7/073048}
}

@article{sarvepalli2012efficient,
  title={Efficient decoding of topological color codes},
  author={Sarvepalli, Pradeep and Raussendorf, Robert},
  journal={Phys. Rev. A},
  volume={85},
  number={2},
  pages={022317},
  year={2012},
  publisher={APS},
  doi={10.1103/PhysRevA.85.022317}
}

@article{bombin2013self,
  title={Self-correcting quantum computers},
  author={Bombin, Hector and Chhajlany, Ravindra W and Horodecki, Micha{\l} and Martin-Delgado, Miguel-Angel},
  journal={New J. Phys.},
  volume={15},
  number={5},
  pages={055023},
  year={2013},
  publisher={IOP Publishing},
  doi={10.1088/1367-2630/15/5/055023}
}

@article{pogorelov2025experimental,
  title={Experimental fault-tolerant code switching},
  author={Pogorelov, Ivan and Butt, Friederike and Postler, Lukas and Marciniak, Christian D and Schindler, Philipp and M{\"u}ller, Markus and Monz, Thomas},
  journal={Nature Physics},
  volume={21},
  number={2},
  pages={298--303},
  year={2025},
  publisher={Nature Publishing Group UK London},
  doi={10.1038/s41567-024-02727-2}
}

@article{postler2022demonstration,
  title={Demonstration of fault-tolerant universal quantum gate operations},
  author={Postler, Lukas and Heu$\beta$en, Sascha and Pogorelov, Ivan and Rispler, Manuel and Feldker, Thomas and Meth, Michael and Marciniak, Christian D and Stricker, Roman and Ringbauer, Martin and Blatt, Rainer and others},
  journal={Nature},
  volume={605},
  number={7911},
  pages={675--680},
  year={2022},
  publisher={Nature Publishing Group UK London},
  doi={10.1038/s41586-022-04721-1}
}

@misc{daguerre2025experimental,
  title={Experimental demonstration of high-fidelity logical magic states from code switching},
  author={Daguerre, Lucas and Blume-Kohout, Robin and Brown, Natalie C and Hayes, David and Kim, Isaac H},
  year={2025},
  note = {{P}reprint at arXiv:2506.14169},
  url={https://doi.org/10.48550/arXiv.2506.14169}
}

@article{lacroix2025scaling,
  title={Scaling and logic in the color code on a superconducting quantum processor},
  author={Lacroix, Nathan and Bourassa, Alexandre and Heras, Francisco JH and Zhang, Lei M and Bausch, Johannes and Senior, Andrew W and Edlich, Thomas and Shutty, Noah and Sivak, Volodymyr and Bengtsson, Andreas and others},
  journal={Nature},
  volume={645},
  pages={614--619},
  year={2025},
  publisher={Nature Publishing Group UK London},
  doi={/10.1038/s41586-025-09061-4}
}

@article{nguyen2021demonstration,
  title={Demonstration of {S}hor encoding on a trapped-ion quantum computer},
  author={Nguyen, Nhung H and Li, Muyuan and Green, Alaina M and Huerta Alderete, Cinthia and Zhu, Yingyue and Zhu, Daiwei and Brown, Kenneth R and Linke, Norbert M},
  journal={Phys. Rev. Appl.},
  volume={16},
  pages={024057},
  year={2021},
  publisher={APS},
  doi={10.1103/PhysRevApplied.16.024057}
}

@misc{dasu2025breaking,
  title={Breaking even with magic: demonstration of a high-fidelity logical non-{C}lifford gate},
  author={Dasu, Shival and Burton, Simon and Mayer, Karl and Amaro, David and Gerber, Justin A and Gilmore, Kevin and Gresh, Dan and DelVento, Davide and Potter, Andrew C and Hayes, David},
  year={2025},
  note = {{P}reprint at arXiv:2506.14688},
  url={https://doi.org/10.48550/arXiv.2506.14688}  
}

@article{bravyi2012magic,
  title={Magic-state distillation with low overhead},
  author={Bravyi, Sergey and Haah, Jeongwan},
  journal={Phys. Rev. A},
  volume={86},
  number={5},
  pages={052329},
  year={2012},
  publisher={APS},
  doi={10.1103/PhysRevA.86.052329}
}

@misc{wang2009graphical,
  title={Graphical algorithms and threshold error rates for the 2{D} colour code},
  author={Wang, David S and Fowler, Austin G and Hill, Charles D and Hollenberg, Lloyd CL},
  note = {{P}reprint at arXiv:0907.1708},
  url={https://doi.org/10.48550/arXiv.0907.1708},
  year={2009}
}

@misc{stephens2014efficient,
  title={Efficient fault-tolerant decoding of topological color codes},
  author={Stephens, Ashley M},
  note = {{P}reprint at arXiv:1402.3037},
  url={https://doi.org/10.48550/arXiv.1402.3037},
  year={2014}
}

@article{chamberland2020topological,
  title={Topological and subsystem codes on low-degree graphs with flag qubits},
  author={Chamberland, Christopher and Zhu, Guanyu and Yoder, Theodore J and Hertzberg, Jared B and Cross, Andrew W},
  journal={Phys. Rev. X},
  volume={10},
  number={1},
  pages={011022},
  year={2020},
  publisher={APS},
  doi={10.1103/PhysRevX.10.011022}
}

@article{lee2025color,
  title={Color code decoder with improved scaling for correcting circuit-level noise},
  author={Lee, Seok-Hyung and Li, Andrew and Bartlett, Stephen D},
  journal={Quantum},
  volume={9},
  pages={1609},
  year={2025},
  publisher={Verein zur F{\"o}rderung des Open Access Publizierens in den Quantenwissenschaften},
  doi={10.22331/q-2025-01-27-1609}
}

@article{Metropolis_1949,
  author = {Nicholas Metropolis and S. Ulam},
  title = {The {M}onte {C}arlo Method},
  journal = {Journal of the American Statistical Association},
  volume = {44},
  number = {247},
  pages = {335--341},
  year = {1949},
  publisher = {ASA Website},
  doi = {10.1080/01621459.1949.10483310},
  note ={PMID: 18139350}
}

@book{Rubino_2009,
  title={Rare event simulation using Monte Carlo methods},
  author={Rubino, Gerardo and Tuffin, Bruno},
  publisher = {John Wiley \& Sons, Ltd},
  isbn = {9780470745403},
  doi = {10.1002/9780470745403},
  year = {2009}
}

@article{higgott2022pymatching,
  title={Pymatching: A python package for decoding quantum codes with minimum-weight perfect matching},
  author={Higgott, Oscar},
  journal={ACM Transactions on Quantum Computing},
  volume={3},
  number={3},
  pages={1--16},
  year={2022},
  publisher={ACM New York, NY},
  doi={10.1145/3505637}
}

@article{gidney2021stim,
  title={Stim: a fast stabilizer circuit simulator},
  author={Gidney, Craig},
  journal={Quantum},
  volume={5},
  pages={497},
  year={2021},
  publisher={Verein zur F{\"o}rderung des Open Access Publizierens in den Quantenwissenschaften},
  doi={10.22331/q-2021-07-06-497}
}

@misc{Landahl_2011,
  title={Fault-tolerant quantum computing with color codes},
  author={Andrew J. Landahl and Jonas T. Anderson and Patrick R. Rice},
  note = {{P}reprint at arXiv:quant-ph/1108.5738},
  year={2011},
  url={https://doi.org/10.48550/arXiv.1108.5738}
}

@article{Li_2017,
  title = {Fault tolerance with bare ancillary qubits for a [[7,1,3]] code},
  author = {Li, Muyuan and Guti\'errez, Mauricio and David, Stanley E. and Hernandez, Alonzo and Brown, Kenneth R.},
  journal = {Phys. Rev. A},
  volume = {96},
  issue = {3},
  pages = {032341},
  numpages = {10},
  year = {2017},
  month = {9},
  publisher = {American Physical Society},
  doi = {10.1103/PhysRevA.96.032341},
  url = {https://link.aps.org/doi/10.1103/PhysRevA.96.032341}
}

@article{heussen2024dynamical,
  title={Dynamical subset sampling of quantum error-correcting protocols},
  author={Heu{\ss}en, Sascha and Winter, Don and Rispler, Manuel and M{\"u}ller, Markus},
  journal={Phys. Rev. Res.},
  volume={6},
  number={1},
  pages={013177},
  year={2024},
  publisher={APS},
  doi={10.1103/PhysRevResearch.6.013177}
}

@article{Sullivan_2019_pyvista,
  doi = {10.21105/joss.01450},
  url = {https://doi.org/10.21105/joss.01450},
  year = {2019},
  month = {5},
  publisher = {The Open Journal},
  volume = {4},
  number = {37},
  pages = {1450},
  author = {C. Bane Sullivan and Alexander Kaszynski},
  title = {{PyVista}: 3{D} plotting and mesh analysis through a streamlined interface for the {V}isualization {T}oolkit ({VTK})},
  journal = {Journal of Open Source Software}
}

@article{vtkBook,
  title={Visualization toolkit software},
  author={Wills, Graham},
  journal={Wiley Interdisciplinary Reviews: Computational Statistics},
  volume={4},
  number={5},
  pages={474--481},
  year={2012},
  publisher={Wiley Online Library},
  doi={10.1002/wics.1224}
}

@article{Bombin_2006,
  title = {Topological Quantum Distillation},
  author = {Bombin, H. and Martin-Delgado, M. A.},
  journal = {Phys. Rev. Lett.},
  volume = {97},
  issue = {18},
  pages = {180501},
  numpages = {4},
  year = {2006},
  month = {10},
  publisher = {American Physical Society},
  doi = {10.1103/PhysRevLett.97.180501},
  url = {https://link.aps.org/doi/10.1103/PhysRevLett.97.180501}
}

@article{Bombin_2007,
  title = {Topological Computation without Braiding},
  author = {Bombin, H. and Martin-Delgado, M. A.},
  journal = {Phys. Rev. Lett.},
  volume = {98},
  issue = {16},
  pages = {160502},
  numpages = {4},
  year = {2007},
  month = {4},
  publisher = {American Physical Society},
  doi = {10.1103/PhysRevLett.98.160502},
  url = {https://link.aps.org/doi/10.1103/PhysRevLett.98.160502}
}

@article{kubica2015unfolding,
  title={Unfolding the color code},
  author={Kubica, Aleksander and Yoshida, Beni and Pastawski, Fernando},
  journal={New J. Phys.},
  volume={17},
  number={8},
  pages={083026},
  year={2015},
  publisher={IOP Publishing},
  doi={10.1088/1367-2630/17/8/083026}
}

@misc{Chamfered_Cubes,
  author = {Weisstein, Eric W.},
  title  = {Chamfered Cube},
  year   = {2024},
  url    = {https://mathworld.wolfram.com/ChamferedCube.html},
  note   = {MathWorld -- A Wolfram Web Resource},
}

@misc{gidney2023new,
  title={New circuits and an open source decoder for the color code},
  author={Gidney, Craig and Jones, Cody},
  note = {{P}reprint at arXiv:2312.08813},
  year={2023},
  url={https://doi.org/10.48550/arXiv.2312.08813}
}

@article{kubica2023efficient,
  title={Efficient color code decoders in $d \geq 2$ dimensions from toric code decoders},
  author={Kubica, Aleksander and Delfosse, Nicolas},
  journal={Quantum},
  volume={7},
  pages={929},
  year={2023},
  publisher={Verein zur F{\"o}rderung des Open Access Publizierens in den Quantenwissenschaften},
  doi={10.22331/q-2023-02-21-929}
}

@misc{turner2020decoder,
  title={A decoder for the color code with boundaries},
  author={Turner, Skylar and Hanish, Josey and Blanchard, Eion and Davis, Noah and La Cour, Brian},
  note = {{P}reprint at arXiv:2003.11602},
  year={2020},
  url={https://doi.org/10.48550/arXiv.2003.11602}
}

@article{delfosse2014decoding,
  title={Decoding color codes by projection onto surface codes},
  author={Delfosse, Nicolas},
  journal={Phys. Rev. A},
  volume={89},
  number={1},
  pages={012317},
  year={2014},
  publisher={APS},
  doi={10.1103/PhysRevA.89.012317}
}

@article{kubica2015universal,
  title={Universal transversal gates with color codes: A simplified approach},
  author={Kubica, Aleksander and Beverland, Michael E},
  journal={Phys. Rev. A},
  volume={91},
  number={3},
  pages={032330},
  year={2015},
  publisher={APS},
  doi={10.1103/PhysRevA.91.032330}
}

@misc{bombin2018transversal,
  title={Transversal gates and error propagation in 3{D} topological codes},
  author={Bombin, Hector},
  note = {{P}reprint at arXiv:1810.09575},
  url={https://doi.org/10.48550/arXiv.1810.09575},
  year={2018},
}

@misc{beni2025tesseract,
  title={Tesseract: A search-based decoder for quantum error correction},
  author={Beni, Laleh Aghababaie and Higgott, Oscar and Shutty, Noah},
  note = {{P}reprint at arXiv:2503.10988},
  url={https://doi.org/10.48550/arXiv.2503.10988},
  year={2025}
}

@article{baireuther2019neural,
  title={Neural network decoder for topological color codes with circuit level noise},
  author={Baireuther, Paul and Caio, Marcello D and Criger, Ben and Beenakker, Carlo WJ and O’Brien, Thomas E},
  journal={New J. Phys.},
  volume={21},
  number={1},
  pages={013003},
  year={2019},
  publisher={IOP Publishing},
  doi={10.1088/1367-2630/aaf29e}
}

@misc{koutsioumpas2025colour,
  title={Colour Codes Reach Surface Code Performance using Vibe Decoding},
  author={Koutsioumpas, Stergios and Noszko, Tamas and Sayginel, Hasan and Webster, Mark and Roffe, Joschka},
  note = {{P}reprint at arXiv:2508.15743},
  url={https://doi.org/10.48550/arXiv.2508.15743},
  year={2025}
}

@article{dennis2002topological,
  title={Topological quantum memory},
  author={Dennis, Eric and Kitaev, Alexei and Landahl, Andrew and Preskill, John},
  journal={J. Math. Phys.},
  volume={43},
  number={9},
  pages={4452--4505},
  year={2002},
  publisher={American Institute of Physics},
  doi={10.1063/1.1499754}
}

@article{chamberland2016thresholds,
  title={Thresholds for universal concatenated quantum codes},
  author={Chamberland, Christopher and Jochym-O’Connor, Tomas and Laflamme, Raymond},
  journal={Phys. Rev. Lett.},
  volume={117},
  number={1},
  pages={010501},
  year={2016},
  publisher={APS},
  doi={10.1103/PhysRevLett.117.010501}
}

@misc{paetznick2011fault,
  title={Fault-tolerant ancilla preparation and noise threshold lower bounds for the 23-qubit Golay code},
  author={Paetznick, Adam and Reichardt, Ben W},
  note = {{P}reprint at arXiv:1106.2190},
  year={2011},
  url={https://doi.org/10.48550/arXiv.1106.2190}
}

@misc{svore2005flow,
  title={A flow-map model for analyzing pseudothresholds in fault-tolerant quantum computing},
  author={Svore, Krysta M and Cross, Andrew W and Chuang, Isaac L and Aho, Alfred V},
  note = {{P}reprint at arXiv:quant-ph/0508176},  
  year={2005},
  url={https://doi.org/10.48550/arXiv.quant-ph/0508176}
}

@article{gupta2024encoding,
  title={Encoding a magic state with beyond break-even fidelity},
  author={Gupta, Riddhi S and Sundaresan, Neereja and Alexander, Thomas and Wood, Christopher J and Merkel, Seth T and Healy, Michael B and Hillenbrand, Marius and Jochym-O’Connor, Tomas and Wootton, James R and Yoder, Theodore J and others},
  journal={Nature},
  volume={625},
  number={7994},
  pages={259--263},
  year={2024},
  publisher={Nature Publishing Group UK London},
  doi={10.1038/s41586-023-06846-3}
}

@misc{bluvstein2025architectural,
  title={Architectural mechanisms of a universal fault-tolerant quantum computer},
  author={Bluvstein, Dolev and Geim, Alexandra A and Li, Sophie H and Evered, Simon J and Ataides, J and Baranes, Gefen and Gu, Andi and Manovitz, Tom and Xu, Muqing and Kalinowski, Marcin and others},
  note = {{P}reprint at arXiv:2506.20661}, 
  year={2025},
  url={https://doi.org/10.48550/arXiv.2506.20661}
}

@article{kubica2018three,
  title={Three-dimensional color code thresholds via statistical-mechanical mapping},
  author={Kubica, Aleksander and Beverland, Michael E and Brand{\~a}o, Fernando and Preskill, John and Svore, Krysta M},
  journal={Phys. Rev. Lett.},
  volume={120},
  number={18},
  pages={180501},
  year={2018},
  publisher={APS},
  doi={10.1103/PhysRevLett.120.180501}
}

@article{Edmonds_1965,
   title={Paths, Trees, and Flowers},
   volume={17}, DOI={10.4153/CJM-1965-045-4},
   journal={Canadian Journal of Mathematics},
   author={Edmonds, Jack},
   year={1965},
   pages={449–467}
}

@article{Dennis_2002,
  author = {Dennis, Eric and Kitaev, Alexei and Landahl, Andrew and Preskill, John},
  title = {Topological quantum memory},
  journal = {J. Math. Phys.},
  volume = {43},
  number = {9},
  pages = {4452-4505},
  year = {2002},
  month = {09},
  issn = {0022-2488},
  doi = {10.1063/1.1499754}
}

@misc{aliferis2005quantum,
  title={Quantum accuracy threshold for concatenated distance-3 codes},
  author={Aliferis, Panos and Gottesman, Daniel and Preskill, John},
  note = {{P}reprint at arXiv:quant-ph/0504218},
  year={2005},
  url={https://doi.org/10.48550/arXiv.quant-ph/0504218}
}

@article{Chamberland_2020,
  doi = {10.1088/1367-2630/ab68fd},
  url = {https://dx.doi.org/10.1088/1367-2630/ab68fd},
  year = {2020},
  month = {2},
  publisher = {IOP Publishing},
  volume = {22},
  number = {2},
  pages = {023019},
  author = {Christopher Chamberland and Aleksander Kubica and Theodore J Yoder and Guanyu Zhu},
  title = {Triangular color codes on trivalent graphs with flag qubits},
  journal = {New J. Phys.},
}

@article{paler2023pipelined,
  title={Pipelined correlated minimum weight perfect matching of the surface code},
  author={Paler, Alexandru and Fowler, Austin G},
  journal={Quantum},
  volume={7},
  pages={1205},
  year={2023},
  publisher={Verein zur F{\"o}rderung des Open Access Publizierens in den Quantenwissenschaften},
  doi={10.22331/q-2023-12-12-1205}
}

@misc{fowler2013optimal,
  title={Optimal complexity correction of correlated errors in the surface code},
  author={Fowler, Austin G},
  year={2023},
  note = {{P}reprint at arXiv:1310.0863},
  url={https://doi.org/10.48550/arXiv.1310.0863}
}

@article{fowler2009high,
  title={High-threshold universal quantum computation on the surface code},
  author={Fowler, Austin G and Stephens, Ashley M and Groszkowski, Peter},
  journal={Phys. Rev. A},
  volume={80},
  number={5},
  pages={052312},
  year={2009},
  publisher={APS},
  doi={10.1103/PhysRevA.80.052312}
}

@article{Kitaev_2003,
  title = {Fault-tolerant quantum computation by anyons},
  journal = {Annals of Physics},
  volume = {303},
  number = {1},
  pages = {2-30},
  year = {2003},
  issn = {0003-4916},
  doi = {https://doi.org/10.1016/S0003-4916(02)00018-0},
  url = {https://www.sciencedirect.com/science/article/pii/S0003491602000180},
  author = {A.Yu. Kitaev},
}

@misc{bravyi2012subsystem,
  title={Subsystem surface codes with three-qubit check operators},
  author={Bravyi, Sergey and Duclos-Cianci, Guillaume and Poulin, David and Suchara, Martin},
  year={2012},
  note = {{P}reprint at arXiv:207.1443},
  url={https://doi.org/10.48550/arXiv.1207.1443}
}

@article{butt2024fault,
  title={Fault-tolerant code-switching protocols for near-term quantum processors},
  author={Butt, Friederike and Heu{\ss}en, Sascha and Rispler, Manuel and M{\"u}ller, Markus},
  journal={PRX Quantum},
  volume={5},
  number={2},
  pages={020345},
  year={2024},
  publisher={APS},
  doi={0.1103/PRXQuantum.5.020345}
}

@article{bombin2016dimensional,
  title={Dimensional jump in quantum error correction},
  author={Bomb{\'\i}n, H{\'e}ctor},
  journal={New J. Phys.},
  volume={18},
  number={4},
  pages={043038},
  year={2016},
  publisher={IOP Publishing},
  doi={0.1088/1367-2630/18/4/043038}
}

@article{anderson2014fault,
  title = {Fault-Tolerant Conversion between the {S}teane and {R}eed-{M}uller Quantum Codes},
  author = {Anderson, Jonas T. and Duclos-Cianci, Guillaume and Poulin, David},
  journal = {Phys. Rev. Lett.},
  volume = {113},
  issue = {8},
  pages = {080501},
  numpages = {5},
  year = {2014},
  month = {Aug},
  publisher = {American Physical Society},
  doi = {10.1103/PhysRevLett.113.080501},
  url = {https://link.aps.org/doi/10.1103/PhysRevLett.113.080501}
}

@article{honciuc2024implementing,
  title={Implementing fault-tolerant non-{C}lifford gates using the [[8, 3, 2]] color code},
  author={Honciuc Menendez, Daniel and Ray, Annie and Vasmer, Michael},
  journal={Phys. Rev. A},
  volume={109},
  number={6},
  pages={062438},
  year={2024},
  publisher={APS},
  doi={0.1103/PhysRevA.109.062438}
}

@misc{senior2025scalable,
  title={A scalable and real-time neural decoder  for topological quantum codes},
  author={Andrew W. Senior and Thomas Edlich and Francisco J. H. Heras and Lei M. Zhang and Oscar Higgott and James S. Spencer and Taylor Applebaum and Sam Blackwell and Justin Ledford and Akvilė Žemgulytė and Augustin Žídek and Noah Shutty and Andrew Cowie and Yin Li and George Holland and Peter Brooks and Charlie Beattie and Michael Newman and Alex Davies and Cody Jones and Sergio Boixo and Hartmut Neven and Pushmeet Kohli and Johannes Bausch},
  note = {{P}reprint at arXiv:quant-ph/2512.07737},  
  year={2025},
  url={https://arxiv.org/abs/2512.07737}
}

\end{document}